\documentclass{acm_proc_article-sp}

\usepackage{graphicx}
\usepackage{epstopdf} 
\usepackage{multirow}

\usepackage{etoolbox}
\makeatletter
\patchcmd{\maketitle}{\@copyrightspace}{}{}{}
\makeatother

\begin{document}

\title{Tactical communication systems based on civil standards:
\\ Modeling in the MiXiM framework}

\numberofauthors{2} 
%
\author{
%
\alignauthor Humberto Escudero Argumanez\\
\alignauthor Matthias Tschauner\\
\and
       \affaddr{Fraunhofer Institute for Communication, Information Processing and Ergonomics (FKIE)}\\
       \affaddr{Communication Systems Department (KOM) - Software Defined Radio Group (SDR)} \\
       \affaddr{Wachtberg, Germany}\\
\and
       \email{\{humberto.escudero.argumanez|matthias.tschauner\}@fkie.fraunhofer.de}
\and  
}
\maketitle

\begin{abstract}

In this paper, new work is presented belonging to an ongoing study\footnote{This research project was performed under contract with the Federal Office of Bundeswehr Equipment, Information Technology and In-Service Support (BAAINBw), Germany.}, which evaluates civil communication standards as potential candidates for the future military Wide Band Waveforms (WBWFs). After an evaluation process of possible candidates presented in \cite{couturier2011evaluation}, the selection process in \cite{liedtke2012selection} showed that the IEEE 802.11n OFDM could be a possible military WBWF candidate, but it should be further investigated first in order to enhance or even replace critical modules. According to this, some critical modules of the physical layer has been further analyzed in \cite{liedtke2013electronic} regarding the susceptibility of the OFDM signal under jammer influences. However, the critical modules of the MAC layer (e.g., probabilistic medium access CSMA/CA) have not been analysed. In fact, it was only suggested in \cite{liedtke2012selection} to replace this medium access by the better suited Unified Slot Allocation Protocol - Multiple Access (USAP-MA)\cite{young1999usapma}.
In this regard, the present contribution describes the design paradigms of the new MAC layer and explains how the proposed WBWF candidate has been modelled within the MiXiM Framework of the OMNeT++ simulator.

\end{abstract}


\section{Introduction}
The international defence forces stride ahead towards further development of the Network Enabled Capabilities (NEC) for the network centric warfare.
In this regard, it has been foreseen that the future tactical wireless communication systems are based on the Software Defined Radio (SDR) technology.
The SDR technology allows to provide different wireless communication capabilities by means of exchanging the software deployed on the SDR device, which is commonly named as Waveform (WF). This term comprises of all layers in the ISO/OSI model.
The military operations planned and executed at the battalion level and below are tactical ones and commonly referred as the ``last mile''.
Due to the heterogeneity of the operations, new and more powerful wireless communication capabilities are demanded. Thus, such capabilities will be provided in form of new WFs.
Generally, the goal of any WF is to fulfil the necessity for Command and Control (C2) and Coordination (C3) services on the move. In order to achieve such goal, a WF should provide for instance adequate communication range, sufficient transfer capacity, the support of IP and Quality of Service (QoS) traffics, dynamic relaying, flooding or routing abilities in order to keep the connectivity on the move, among others. Because of the military nature of the communication deployments, the WF will have to face communication outage given for instance by the mobility, changes in the physical surroundings and jamming situations, etc. Thus, smarter protocols and algorithms should be designed for the new WFs in order to overcome such outage. This basically will help to create self-organizing WFs that react and adapt to the environment changes while providing connectivity.

The common deployed WF in the last mile is the so-called Combat Network Radio Narrow Band WF (CNR-NBWF). This NBWF cannot cover all new demanded capabilities due to its narrow bandwidth, i.e., < 100 kHz.
Due to this, military efforts placed the focus on new military Wide Band WFs (WBWFs), which require higher spectral bandwidths, i.e., $\geq$ 1 MHz, and provide data rates > 1 Mbps. Thus, efforts from international organizations and consortia are pursued in this area. Furthermore, several nations are aiming to design additional WBWFs for national use.  
This necessity for investigating new possible WBWFs has increased the research in the military communications area. Hence, the preceding phase of our work studied the suitability of the current civil standards for wireless communications in the military domain. 
For instance, the work in \cite{couturier2011evaluation} sketched the fundamentals of the tactical communications, which served as pillar in the evaluation of civil communication systems. In essence, four different types of civil standards (i.e., broadcast, cellular, wireless local networks and trunked radio systems) were studied and rated according to the operational and technical military requirements. Subsequently, the work in \cite{liedtke2012selection} conducted the selection of possible candidates according to the following criteria: modern WBWF systems, bi-directionality and scalability.
The outcome gave the following three Orthogonal Frequency Division Multiplex (OFDM) candidates: 4th Generation Long Term Evolution (4G/LTE), IEEE 802.16e (WiMaX) and the IEEE 802.11n (WLAN). 
In order to contribute to discrimination, more pragmatic factors were taken as well into account in the final assessment. For the selected system, IEEE 802.11n, the first critical modules were identified and revised. At the PHY layer, the preamble coding and its detection were revised and a longer cyclic prefix was suggested for the purpose to reach a larger communication range. Furthermore, the work in  \cite{liedtke2013electronic} exhibited the susceptibility of the OFDM signal under jamming situations and hence, a novel anti-jam demodulator was provided for mitigating such jammer influences.
The critical module identified at the Medium Access Control (MAC) layer was the probabilistic medium access. So, instead of the originally included Carrier Sensed Multiple Access with Collision Avoidance (CSMA/CA), the more suited Unified Slot Allocation Protocol Multiple Access (USAP-MA) \cite{young1999usapma} was suggested to be used. However, no further related analysis has been done since then. According to this, our present contribution introduces the basics for conducting such task. More precisely, in Section \ref{sec_macdesign} the MAC layer design is outlined. It is then explained for instance how the suggested USAP-MA has been modified and optimized with the inclusion of enhancement techniques provided by the IEEE 802.11n MAC. 
In Section \ref{sec_miximodel} it is laid out how the WBWF has been modelled within the MiXiM framework of the OMNeT++ simulator. Furthermore, the modelling considerations of all currently  programmed layers are explained.
In order to extend our investigations, it is required to consider the overall behaviour of the proposed military WBWF in realistic tactical scenarios. For this, Section \ref{future} presents a briefing of our future planned work. At the end, the conclusion are presented.

\section{MAC Layer Design}\label{sec_macdesign}
In the design of our proposed MAC layer, we have taken into account two main protocols. On the one hand is the IEEE 802.11n MAC layer \cite{ieee201280211}, which includes MAC enhancements for achieving higher throughput such as frame aggregation mechanisms. On the other hand is the USAP-MA \cite{young1999usapma} protocol, which is based on a Time Division Multiple Access (TDMA) mechanism and was designed for a tactical military system called Soldier Phone \cite{bittel1999soldier}. Furthermore, the original USAP protocol was presented in \cite{young1996usap}, and the USAP-MA was further studied in \cite{young2000mba} and at latest optimized in \cite{young2006mdl} for a Wide Band Networking Waveform (WNW). 
These two protocols serve to design features in our MAC layer that will aid to multiplex as efficiently as possible two different traffic flows, best-effort traffic such as Internet Protocol version 4 and 6 (IPv4/6) and real-time traffic like voice.

\subsection{TDMA Scheme}
The TDMA scheme is in essence the key part for the multiplexing, and thus, the trickiest part in our MAC layer to be designed due to design rules and challenges, which are later explained. As a basis, we took the TDMA format presented by USAP-MA, which uses three frame types, i.e., management (MGMT), real time data (RT) and best-effort data (BE) frames (Please note that the original USAP names for the slot types are bootstrap, unicast and broadcast). Each frame type is compounded by a certain amount of slots.

The MGMT slots are used for sending what USAP names as Net Manager Operational Packet (NMOP). Essentially, the NMOP is a bit map that reflects the usage of all data slots in the network with a code mapped in a 2-bit entry per slot. Hence, the first challenge of our design arises and the first design rule can be defined as follows: a MGMT slot must be sufficiently big to signalize all data slots in the network. As a remark, the total amount of MGMT slots reflects the maximum amount of nodes expected to be part of the network and the data slots are the slots used for the transmission of BE and RT traffic.

The RT slots have been designed to transfer voice traffic coded with the Mixed-Excitation Linear Predictive Enhancement (MELPe)\cite{melp1995} voice codec. This codec is widely used in the military tactical communications and can provide data rates of 600, 1200 and 2400 bps. 
In addition, the maximum permitted end-to-end latency in tactical communications (typically between 250 ms and 500 ms) presents an interdependency between the RT slot size and the TDMA frame time length. Thus, the second challenge arises and a new design rule can be defined: a RT slot must transfer as many voice coded frames as generated during one TDMA frame. This asserts that the voice communication remains within the Quality of Service (QoS) boundaries. As a remark, our design has taken the data rate of 2400 bps, in which a voice coded frame of 54 bits is generated each 22.5 ms.

The BE slots have been designed according to the IP Maximum Transfer Unit minimum value (MTU-min). As explained in \cite{rfc791} the IPv4 MTU-min is 576 bytes, which is the datagram size that every internet destination must be able to receive either in one piece or in fragments to be reassembled. As stated in \cite{rfc2460} the IPv6 MTU-min is 1280 bytes, hence, on any link that cannot convey a 1280-octet datagram in one piece, link-specific fragmentation and reassembly must be provided at a layer below IPv6. Additionally to this, the IPv6 can contain several extension headers, which are multiple of 8 octets long each. 
Thus, in order to find a trade-off solution, the following design rule is set: payload transferred in the BE slot must be a multiple of 8 bytes. This rule should aid in the fragmentation process of IPv6 datagrams, in case it can be required.
Besides the aforementioned rules, there is an additional challenge, which is the avoidance of time gaps between slots in order to maximize the channel utilization. This can be achieved if the length of each time slot can be divided by an exact number of OFDM symbols.

For designing such a TDMA frame, we have programmed an iterative method that computes the slot time lengths according to different TDMA frame lengths, in which, for each frame length, the weights of MGMT, RT, BE frames are also varied, i.e., the amount of slots per frame. The iterative method uses the time slot computation presented in \cite{liedtke2012selection} with a physical bit rate of 1.625 Mbps and a MAC Header of 176 bits (explained later in Subsection \ref{ss_macpci}). This physical rate corresponds to the High-Throughput (HT) mode of the IEEE 802.11n standard, using 10 MHz bandwidth with the lowest modulation and coding scheme (MCS), i.e., BSPK and 1/2 convolutional rate. With this input, the iterative method provided only three solutions that conformed the above mentioned rules. The solutions are presented in Table \ref{table:tdmaconfig}. 
We understood that the Solution 3 was the optimum one due to the several reasons. For instance, having more MGMT slots within one TDMA frame provides more feedback during Time Slot Allocation (TSA). Thus, the probability of TSA collisions is lower. Another reason is the RT frame, which provides a very good ratio between the maximum amount nodes in the network and RT slots. This means that up to 5 nodes (i.e., 148 nodes divided 36 RT slots) can compete for a RT slot. 
With regard to the BE frame, the amount of BE slots are approximately twice less than in the other solutions, but the slot size allows to transmit 3.6 times more payload. Furthermore, the payload size fits with the IPv4 MTU-min (576 bytes) and is a multiple of 8-octets, which is suitable in case any IPv6 datagram needs to be fragmented at the Logical Link Control (LLC) layer.

\begin{table}
\centering
\caption{TDMA Frame configurations}
\begin{tabular}{|c|c|c|c|} \hline
 & Solution 1& Solution 2& Solution 3\\ \hline
\begin{tabular}[c]{@{}c@{}} TDMA Frame\\time length\end{tabular}&  80 ms&80 ms&128 ms\\ \hline

\multirow{3}{*}{\begin{tabular}[c]{@{}c@{}}MGMT\\Frame\end{tabular}} &  36\%&36\%&37\%\\
 &  28.8 ms&28.8 ms&47.36 ms\\
 &  90 slots&90 slots&148 slots\\ \hline
\multirow{3}{*}{MGMT Slot} &  32 $\mu s$&32 $\mu s$&32 $\mu s$\\
 &  296 bits& 306 bits&290 bits\\
 &  \begin{tabular}[c]{@{}c@{}} 16 padded\\bits\end{tabular}&\begin{tabular}[c]{@{}c@{}} 6 padded\\bits\end{tabular}&\begin{tabular}[c]{@{}c@{}} 22 padded\\bits\end{tabular}\\ \hline 

\multirow{3}{*}{\begin{tabular}[c]{@{}c@{}}Real-Time\\Frame\end{tabular}} &  7.68\%&11.52\%&12.6\%\\
 &  6.144 ms&9.216 ms&16.128 ms\\
 &  16 slots&24 slots&36 slots\\ \hline
\multirow{2}{*}{RT Slot} &  384 $\mu s$&384 $\mu s$&448 $\mu s$\\
 &\begin{tabular}[c]{@{}c@{}} 4 Voice\\Frames \end{tabular}& \begin{tabular}[c]{@{}c@{}} 4 Voice\\Frames \end{tabular}& \begin{tabular}[c]{@{}c@{}} 6 Voice\\Frames \end{tabular}\\ \hline 

\multirow{3}{*}{\begin{tabular}[c]{@{}c@{}}Best-Effort\\Frame\end{tabular}} &  56.32\%&52.48\%&50.4\%\\
 &  45.046 ms&41.984 ms&64.512 ms\\
 &  44 slots&41 slots&21 slots\\ \hline
\multirow{2}{*}{BE Slot} &  1.024 ms&1.024 ms&3.072 ms\\
 &  160 bytes &160 bytes&576 bytes\\ \hline

\hline\end{tabular}
\label{table:tdmaconfig}
\end{table}

\subsection{MAC PDU Format}\label{ss_macpci}
\begin{figure}[h]
 \centering
 \includegraphics[width=.45\textwidth]{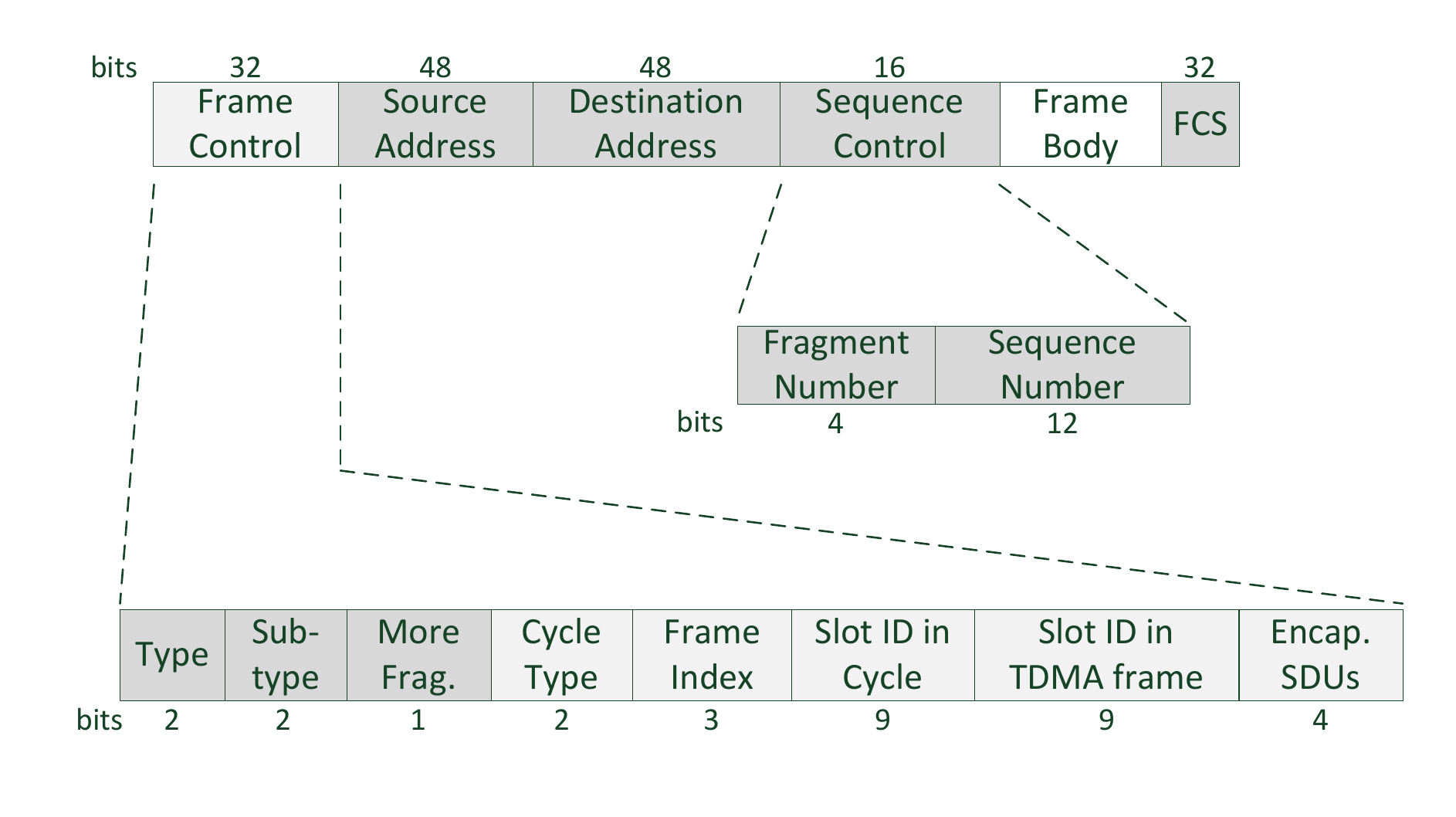} 
 \caption{MAC Protocol Data Unit Format}
 \label{fig:MACPCI}
\end{figure}

The MAC Protocol Data Unit (PDU) format has been taken from the standard IEEE 802.11 \cite{ieee201280211}. It is compounded by a Protocol Control Information (PCI) or Header and Service Data Unit (SDU) or payload. The original MAC PCI has been modified accordingly in order to work with the TDMA frame\footnote{Note that there are two different meanings using the word ``frame''. One kind is related to USAP-MA nomenclature and it is related to the time configuration of the TDMA scheme. The other is related to the OSI Model nomenclature and refers to the MAC Protocol Data Unit}. More precisely, the following original fields have not been further used: \textit{Duration/ID, Address 3, Address 4 and QoS Control}. Regarding the Frame Control field, only the original \textit{Type, Subtype and More Fragment} sub-fields have been used, and new ones have been added. The resulting MAC PDU structure is presented in Figure \ref{fig:MACPCI}, in which the original fields and sub-fields from \cite{ieee201280211} have a darker background.
The resulting MAC PCI uses 176 bits for performing several tasks. For instance, the signalling of sender and receiver is done with 6-octets address blocks. The error-detection is performed as stated in \cite{ieee201280211} with the \textit{Frame Check Sequence} (FCS) field, i.e., a 32-bit cyclic redundancy checksum (CRC) word. The \textit{Sequence Control} field of 16 bits helps to detect duplicated MAC PDUs and identify the fragment sequence of a fragmented PDU.

At this point it is important to remark that according to the USAP-MA, the frame types (i.e., MGMT, RT and BE) contained in a TDMA Frame (also named as \textit{Atomic TDMA Frame}) can be increased, and the group of the same type form a cycle. For instance, if a network should contain 250 nodes, according to our Solution 3 presented in Table \ref{table:tdmaconfig}, the MGMT cycle needs two MGMT frames. In this case, this MGMT cycle is able to allocate up to 296 nodes and has a latency of two Atomic TDMA Frames (i.e., it starts over each 256 ms). The main modification of the MAC PCI has been given within the Frame Control field. While using a TDMA scheme, it is required to index the slots and frames during the network lifetime. For that, the new following sub-fields have been added: \textit{Cycle Type, Frame Index, Slot ID in Cycle and Slot ID in TDMA Frame}. Cycle Type sub-field identifies whether the type is MGMT, RT or BE. In case a cycle is formed by more than 1 frame, Frame Index sub-field indicates which is the position of the current frame in the cycle. This sub-field can index a maximum of 8 frames in a cycle. The slot is indexed twice, i.e., within the cycle and in the Atomic TDMA Frame. The slot index's constraint is the same for both because both sub-fields have 9 bits, hence, no more than 512 slots can be indexed.

Apart from the TDMA configuration and as it has been explained in the beginning of this section, the MAC Layer performance has been boosted with the inclusion of a frame aggregation technique, which is the Multi-SDU aggregation. This technique encapsulates in a single MAC PDU several SDUs belonging to the same traffic and addressed to the same destination. According to \cite{ieee201280211}, this technique should be implemented together with the Block Acknowledgement (Block-ACK) mechanism, however, since our MAC layer uses a TDMA medium access, its usage is not relevant. Regarding the SDUs, the amount encapsulated is indicated by the sub-field \textit{Encapsulated SDUs}. By design, the proposed TDMA configuration shown in Table \ref{table:tdmaconfig} indicates the maximum voice coded frames (or MAC SDUs) to be encapsulated in a RT slot when the MELPe 2.4 kbps voice codec is used. This technique is of special interest for the tactical voice communications because it helps to keep an acceptable QoS.
Depending on the payload or MAC SDU, the MAC PDU can be of different types. The sub-field Type  signalizes this. In our design, the MAC PDU can be a MGMT frame, RT data frame, BE data frame and Push-To-Talk Signalling (PTT-SIG) frame. However, for each MAC PDU type, there might be several sub-types. More details about sub-types are given in the following sub-subsections.

\subsubsection{MGMT PDUs}\label{sss_macmgmt}
The MGMT PDU is exclusively being sent in the MGMT slot. As aforementioned it consists essentially  of a bit map. This bit map contains a 2 bit entry for each data slot (RT and BE), which reflects the slot usage (i.e., idle, transmitting, neighbour transmitting, collision) seen by a node from its local perspective. It is important to remark, that the MGMT PDU has a local impact, thus no global information about the network is included within the bit map. The locally perceived information collected at each neighbour enforces an implicit organization in the overall network. This means for instance that after gathering several bit maps from the neighbourhood, a node is able to identify the hidden node situation, and thus, avoiding interferences and collisions in the future Time Slot Allocations (TSAs).

Each node participating in the network is sending constantly a MGMT PDU with the bit map. This behaviour (called as network-beat) helps to identify the locally perceived neighbourhood and the changes in the network topology, by means of analysing the  link quality and stability of the neighbours. Furthermore, this gathered information is in fact the best feedback to be provided to IP routing/flooding protocols in order to facilitate them the decision making process.
Apart from that, the bit map is also used for conducting a TSA. According to the USAP-MA there are two ways for allocating a time slot, confirmed and unconfirmed. We opted for the unconfirmed mode in order to reduce signalling and speed up the TSA. More details about the slot allocation can be found in \cite{young1999usapma, young1996usap}.

As it has been presented in Table \ref{table:tdmaconfig}, Solution 3 contains 22 padded bits in the MGMT PDU. These padded bits can be used to expand the MGMT functionality by piggybacking responses and notifications. In this sense, for our MGMT PDUs we distinguish three sub-types, i.e., simple beacon, PTT responses (PTT-Res) and Queue Load Level (QLL) notification.

The Beacon is just a MGMT PDU with the bit map. The PTT-Res is piggybacked within the MGMT PDU in order to speed up the PTT session establishment phase. It requires only 16 bits out of the 22 bits. The Most Significant Bit (MSB) is the response (1=positive, 0=negative) and the remaining 15 bits correspond to the PTT session ID. 

The main task of the QLL notification is to influence the BE-Queue Scheduler. This scheduler manages the prioritized BE-Queues by means of a Multilevel Precedence and Preemption (MLPP) scheme. In essence the QLL notification indicates the status of the BE-Queues. Thus, it should aid to grant the access to the resources (i.e., BE slots), in case the network might suffer from congestion. The coding of the QLL notification has not been yet specified, because the BE traffic is not yet enabled.

\subsubsection{PTT-SIG \& RT-DATA PDUs}\label{sss_PTTSIGpdu}

We assume that PTT communication remains to be the most important service to be provided by the WBWFs in the last mile. Thus, due to multi-hop characteristic of the tactical scenarios, the set up of a reliable voice communication is highly desired. Hence, the interferences need to be avoided by means of the following three phases, session establishment, speech and release. In order to do that, instead of using a protocol family like H.323 for Voice over IP (VoIP), we have set the Push-To-Talk Signalling (PTT-SIG) as a type of MAC PDU. This ensures that no additional protocol influences the behaviour of our MAC layer.

The PTT-SIG subtypes are until now only three, i.e., Session Request, Session Release, Session Relay. The exchange of PDUs is very simple. Whenever a new PTT Session needs to be established, the MAC layer receives the correspondent service primitive from the upper layer asking for it. Hence, the MAC layer changes its state in the main FSM in order to reserve a RT slot. 

In this regard, during the \textit{establishment phase} the first use of the RT slot is for a PTT-SIG Session Request PDU, which contains the following information: PTT session ID, voice codec, coding rate and destination address (which can be broadcast, multicast or unicast). For this information, this PDU requires no more than 3 bytes in the payload. For unicast and multicast  PTT Sessions, the responses are required, and as previously explained, these are piggybacked within the MGMT PDUs as a sub-type. This design ensures that the PTT session establishment phase lasts no more than one Atomic TDMA Frame (i.e., according to our design the MGMT cycle is compounded by one MGMT frame). 

When the establishment is successfully completed, the event is signalized to the upper layers and the PTT Session enters in the \textit{speech phase}. Hence, the voice codec starts coding voice frames. These frames are being queued by the MAC layer and will wait until they can be dispatched in the correspondent RT slot. As shown in Table \ref{table:tdmaconfig}, up to 6 MELPe frames of 54 bits can be encapsulated during a RT slot. The encapsulation of these frames is done in a MAC RT-DATA PDU, which is a simple MAC PDU with the type set to RT data and the sub-type identifying the voice codec and coding rate. It is important to remark that the PTT Session ID is not included in the MAC RT-DATA PDU because the filtering at the receiver will be done according to the transmitter.

The \textit{release phase} is performed when the PTT-button is released at the transmitter. In this case, the voice codec stops coding frames and the PTT-SIG Session Release PDU is sent in the next RT slot. After that, all resources (RT slot, memory, etc.) at the transmitter and receiver/s are freed.

If for instance, a transmitter wants to have a PTT session with a node 2-hops away, the PTT Session should be relayed. For that we have included in our design the PTT-SIG Session Relay PDU, whose payload contains all relevant information. As a remark, the transmitter should know prior requesting a PTT Session, to whom it wants to talk and how it can be reached, i.e., who should react as relay. This information could be for instance exchanged through Cross-Layer Optimization (CLO) databases. By now, we have not a final design of this PDU. Thus, no more details about its content can be provided.

\subsubsection{MAC BE-DATA PDU}\label{sss_BEpdu}
This PDU encapsulates the BE traffic, and hence, there is no need for additional signalling or fields.

\section{MiXiM Models}\label{sec_miximodel}
Our work deals as well with the modelling considerations of military tactical systems, which are mainly used in outdoor environments and rely on a high demand in Quality of Service (QoS). 
Thus, the challenges of this work are to create a model as realistic as possible in order to provide good insights into the expected performance according to realistic military tactical scenarios. 
The selected modelling tool was the MiXiM framework \cite{dkopke2008simulating} due to its \textit{Mapping} data-type. 
Regrettably, MiXiM provides models, which have been modelled primarily for indoor and for sensor networks. Hence, new models need to be programmed from scratch for the WBWF modelling. 
In this regard, the present contribution provides the following new models: PHY layer, which is based on the OFDM transmission technique and optimized for outdoor environments; a MAC layer using a TDMA scheme, and a tactical voice communication application based on the PTT behaviour. The details of each model are described in the correspondent subsections.

\subsection{PHY Layer}

A critical issue regarding the accuracy of the PHY layer modelling is the packet-domain level, in which MiXiM works with its \textit{Mapping} data-type. As explained in \cite{wehrle2010modeling}, the relation between channel quality and uncoded Bit Error Rate (BER) is usually non-linear. Therefore, the best to be done during the PHY frame reception is to compute the uncoded BER weighted over time. However, the correctness of this approach has a direct dependency on the following parts: modulation type, Forward Error Correction (FEC) technique and interleaving scheme. This is the reason why the critical modules identified at the PHY layer in \cite{liedtke2012selection} could not be analysed with MiXiM. Bearing in mind the packet-domain, our proposed solution models the PHY layer in such a way that it is able to characterize an outdoor environment without introducing too much complexity in the WBWF modelling. This implies naturally a lack of realism at the PHY layer, but it helps to analyse the performance of upper protocols.

\begin{figure}[t]
 \centering
 \includegraphics[width=0.47\textwidth]{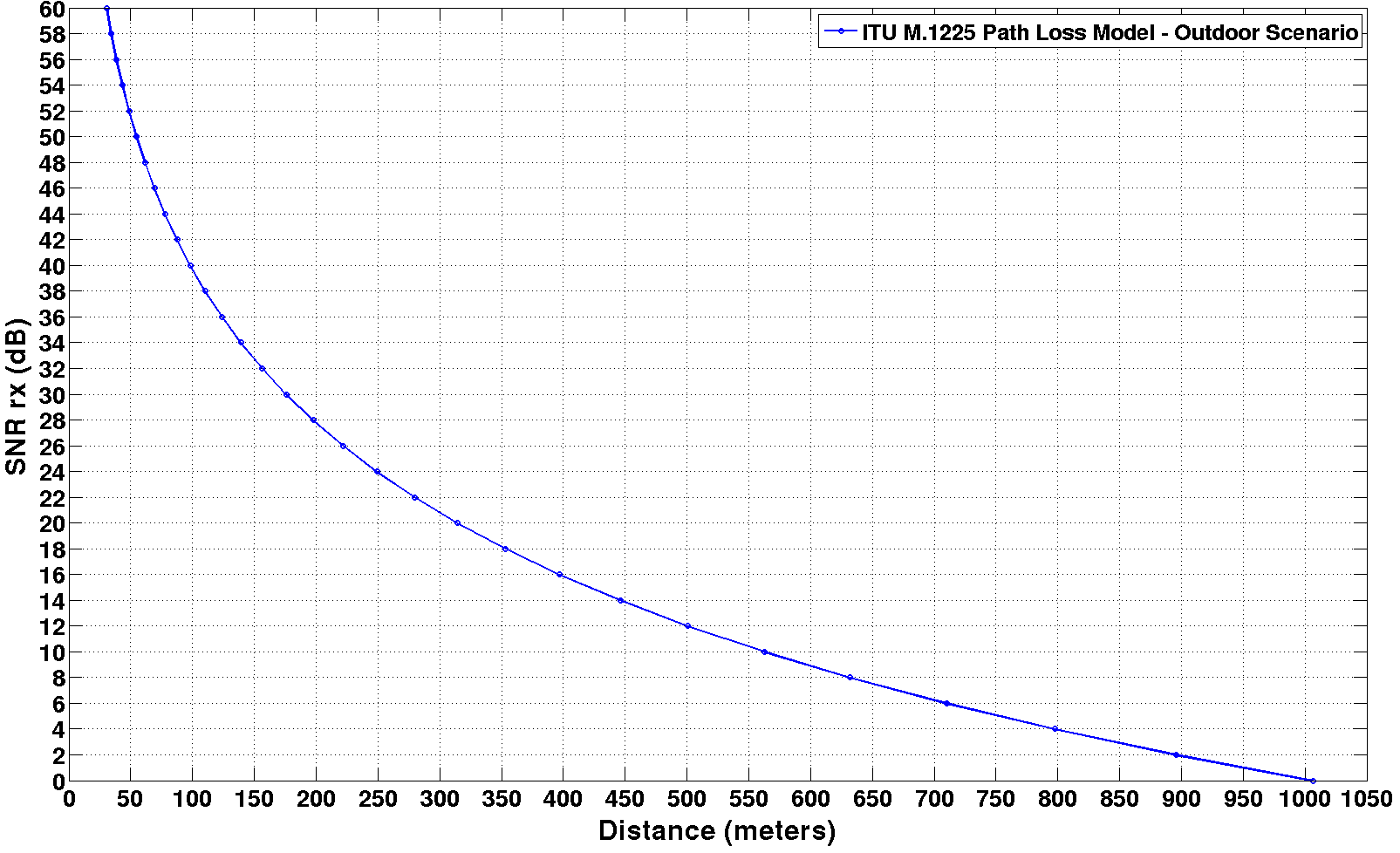} 
 \caption{Vehicular Channel - Path Loss Model}
 \label{fig:plmodel}
\end{figure}

\begin{figure}[t]
 \centering
 \includegraphics[width=0.47\textwidth]{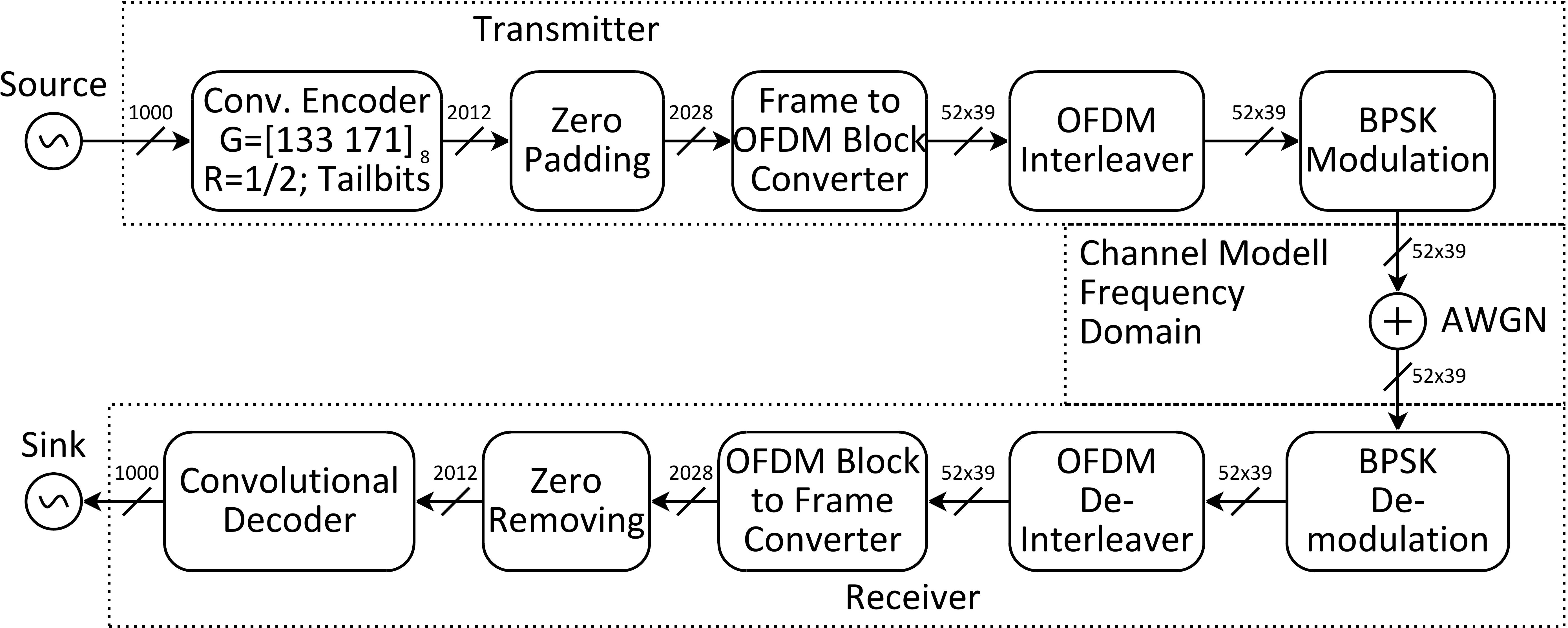} 
 \caption{Block Diagram of the PHY Layer}
 \label{fig:physim}
\end{figure}

Our \textit{PHY Layer} Class has been programmed based on the IEEE 802.11a/n standards. Hence, it considers several bandwidths (BW),i.e., 20, 10, 5, 2.5, 1.25 MHz, and regarding the BW the OFDM symbol and bit rates are adapted. Our layer considers as well different preamble types (i.e., Legacy and Greenfield), the use of 48 or 52 data sub-carriers (SCs), transmission channels in the ISM and UNII bands. For our studies we stick to the High-Throughput (HT) mode (52 data SCs) with Greenfield preamble and a transmit power of 5 watts. The characterization of an outdoor environment has been given by the introduction of a path loss model presented in the ITU Reference M-1225 \cite{mr1225}, more precisely, the Vehicular Channel has been used. This has been programmed as an \textit{Analogue Model} Class and its behaviour is presented in Figure \ref{fig:plmodel}.
The correctness of our approach has been high-tailored to the results obtained from a real PHY Layer implementation, whose block diagram is presented in Figure \ref{fig:physim}. The coded BER results have been used as look-up tables in our \textit{Decider} Class, which is in charge of computing the Packet Error Probability (PER) by means of the Bernoulli Distribution. As indicated in \cite{wehrle2010modeling}, the final success in the reception process is given after a validation with a standard uniformly distributed random number generator.

\subsection{MAC Layer}\label{mac}
Our MAC Layer has been implemented in several Finite State Machines (FSM) and timers. For instance, in order to control the TDMA scheme, the TDMA configuration (Solution 3 in Table \ref{table:tdmaconfig}), loaded in the MAC NEtwork Description (NED) file, is used to schedule a timer. More precisely, each single node in the network is in charge of understanding when the next slot is going to start and how long it lasts. This manner sets the basis for analysing the clock-drift effect in the near future. Whenever a slot is identified to be as ``own'', some other actions can be driven, e.g., transmission of a MAC MGMT PDU in the assigned MGMT slot. The \textit{main FSM} controls basically the MAC layer state according to the TDMA scheme. In this regard, the MAC layer considers whether it is idle, transmitting, receiving, sleeping, searching for a RT slot or BE slot or both slots. In our design, we have considered a fixed assignation of MGMT slots among the nodes. This consideration spares the challenge of how to get a MGMT slot and its consequences. 

Before sending the bitmap at the correspondent MGMT slot, the MAC layer analyses all previously collected bitmaps. Thus, it is able to understand whether TSA collisions occur, or there are hidden nodes. The possible hidden node case is considered prior performing a TSA selection. However, if a TSA has been already selected and notified to the network, it might happen that a collision is reported from another node. Hence, this case is filtered and, if possible, the use of the RT is cancelled. For any reported collision in an own TSA, a node is forced to select a new TSA and delay its transmission. However, this is valid only if the collision happens during the allocation phase and not during the use of the slot. A drawback of the bit map is the lack of resolution. Thus no further details about the hidden nodes can be provided (e.g., address). This could be done by other means like reading the routing tables or performing traffic surveillance. By now, these two ways have not been neither considered nor implemented.

The control of the PTT-SIG communication is also driven in combination with the main MAC FSM. Thus, the stages in the RT slot usage are controlled and the piggybacked PTT Responses are also collected and analysed. The PTT-SIG is being triggered by the service primitives being sent from the PTT layer, which is later explained in Subsection \ref{ptt}. In this regard, the MAC and PTT layers are exchanging service primitive through the control gate of the NED object. This helps to coordinate the triggering, use and release of the PTT session. By now, our MAC layer is able to establish a broadcast and unicast PTT session only for 1-hop neighbours. Due to current stage of our work, we have not been able to complete the implementation of the PTT-SIG, thus, the multicast PTT communication and relay capabilities are missing. Important issues in this regard are the address mapping to be used for the PTT-SIG PDUs and the future unavoidable modifications to be done after this model is integrated with the INET framework.

\subsection{PTT Layer}\label{ptt}
The modelling of a Push-To-Talk (PTT) behaviour is a challenging task that requires, whenever is possible, real input data. Work towards this direction has been already performed for disaster area scenarios as presented in \cite{aschenbruck2006ptt}. To the best of our knowledge we have not found any publication about such measurements results for military environments. Thus, we have not been able to create a speaker model that reflects a realistic behaviour pattern. At this point, it is important to remark that such measurements are dependant of a moving pattern behaviour, which should be as well recorded, analysed and if possible modelled analytically. Such ``future'' model should determine the necessity to talk according to the circumstances of the tactical operational scenario.
Bearing in mind these difficulties, our proposed model contains just a state model that models the events when a button is pressed, released and when the user talks. The logic for the communication with the MAC layer in order to ask for services is also implemented in form of \textit{Control Information} Class.

\section{Future Work}\label{future}
The work towards the MAC layer modelling and the upper layers is still not finished. For instance, the current MAC layer requires more evaluation tests in order to assert that there are no singularities in our code. Besides the completion of the MAC layer, the imminent phase of our work will deal with the inclusion of the INET framework. For that, the Network Interface Card (NIC) NEtwork Description (NED) object require more gates, which will be connected to the INET modules. This change, implies additional modifications in our MAC layer code, for instance, the addressing at the MAC layer needs to be updated. 
If this stage can be finished, it is also planned to extend the information collected at the MAC layer into an additional data base module, which is supposed to track all statistics regarding the behaviour of the wireless medium. This module can be used as an information source for boosting the performance of algorithms implemented at network layer (e.g., path discovery, minimum connecting set discovery, routing, flooding, among others). The realism of the wireless channel model can be increased with more specific channel propagation models like the ones presented in \cite{fischer2012ploss, fischer2012shadowing, fischer2013tapdelay}. These, should be combined with a mobility model that allows to match the outage caused by the speed and physical surroundings.

\section{Conclusions}\label{conclusions}
This paper has presented the work towards design and modelling considerations of a future military WBWF, which is based on a civil communication standard, i.e., IEEE 802.11n. 
The design of a new MAC layer, which includes the TDMA scheme provided by the Unified Slot Allocation Protocol - Multiple Access (USAP-MA)\cite{young1999usapma}, has been explained. Furthermore, detailed description about how the TDMA frame format has been conceived for our WBWF and how the original MAC Protocol Data Unit (PDU) Format from \cite{ieee201280211} has been accordingly modified has been given. Subsequently, the WBWF modelling in the MiXiM framework of the OMNeT++ simulator has been explained. In this regard, the need to include the INET framework as part of our modelling has been clearly detailed. As a final remark, the work to be complete as well as the work to be conducted in the near future has been also also presented.

%
\bibliographystyle{unsrt}
\bibliography{references}  
\end{document}